# Tackling Erosion in Variant-Rich Software Systems: Challenges and Approaches

Johannes Stümpfle[a,]*, Nasser Jazdi[a], Michael Weyrich[a]

[a]*Institute of Industrial Automation and Software Engineering, University of Stuttgart, Germany*

* Corresponding author. Tel.: +49 711 685 67295; *E-mail address:* johannes.stuempfle@ias.uni-stuttgart.de

**Abstract**

Software product lines (SPL) have emerged as a pivotal paradigm in software engineering, enabling the efficient development of variant-rich software systems. Consistently updating these systems, often through over-the-air updates, enables the continuous integration of new features and bug fixes, ensuring the system remains up to date throughout its entire lifecycle. However, evolving such complex systems is an error prone task, leading to a phenomenon known as erosion. This phenomenon significantly impacts the efficiency and longevity of software systems, presenting a formidable challenge for manufacturers of variant-rich software systems, such as in the automotive domain. While existing studies concentrate on the evolutionary planning of variant-rich software systems, there is a noticeable lack of research addressing the problem of erosion. In this paper, we conduct an in-depth exploration of the erosion phenomena within variant-rich software systems. We begin by highlighting the significance of controlling erosion in extensive variant-rich software systems. Subsequently, we address the current challenges regarding tackling erosion, including issues such as the lack of a consensus on understanding and defining erosion, as well as the early detection and elimination. Finally, we outline a first approach aimed at tackling erosion in variant-rich software systems.

*Keywords:* Software Product Line; Erosion; Design: challenges & Innovation

## 1. Introduction

In today's rapidly evolving technological landscape, software-intensive systems have become the backbone of numerous industries. A prime example is the automotive industry, undergoing the complex transition from mechanical-defined to software-defined vehicles [1]. By serving a big variety of customer requirements and market regulations, the automotive industry is faced with the challenge of developing and managing a wide range of vehicle variants (variability in space). With the goal of continuously updating the software to add new features or fixing bugs a second variability is introduced (variability in time) [2]. When modifying, enhancing, and adapting such variant-rich software systems (VRS) over time, erosion emerges as a critical issue.

Software erosion in general describes the gradually deviation of a software system from its intended design and functionality. The concept is used in a broad context, whereby the phenomenon of erosion can be viewed from different perspectives. Thus, the causes of erosion are much more far-reaching than just the evolution of software systems. [3]

In this paper, we focus on the evolution of VRS, due to the motivation of continuously updating software in VRS. Thus, we conceptualize erosion in VRS as a deterioration of the system over time, due to its complexity arising from the multitude of variants. This complexity encompasses causative technical factors such as the accumulation of technical dept as well as non-technical factors such as documentation gaps. Our examination extends beyond architectural levels, also including erosion at the code level, all resulting in consequences such as quality (maintainability, evolvability, efficiency and usability) degradation, poor system integrity and shortened system lifetime. Thus, hindering the performance and continuous evolution of variant-rich software systems.

The impact of software erosion in variant-rich software systems is notably evident in the automotive domain. As these systems evolve, the intricate interplay of diverse software variants introduces challenges that extend beyond traditional mechanical considerations. This impact manifests in the form of heightened complexity [4], affecting not only the functionality but also the safety and reliability of vehicles.

Thus, the goal of this paper is to provide an overview in the field of erosion in variant-rich software systems based on the systems evolution and the associated challenges.

The remainder of this paper is structured as follows: In chapter two we start by showing the basics of software product lines as enablers of developing variant-rich software systems, their evolution, and the consequential erosion. Chapter 3 outlines five key challenges we derived in tackling erosion in VRS. We present our first approach for tackling erosion in chapter four, based on the derived challenges. In the final chapter five we give an outlook regarding future work in this field and further research directions.

**2. Basic and Related Work**

In this chapter we present the main background knowledge about variant-rich software systems focusing on software product lines and their evolution.

*2.1. Software Product Lines*

Developing variant-rich software systems is a complex and error-prone task, considering the vast amount of different product variants and its dependencies. A generic programming approach for developing variant-rich software systems are software product lines. They base on a systematic reuse of software artifacts to create a family of related products efficiently. Instead of building each software product from scratch, SPL leverage a common core of reusable components, modules etc. tailored to meet the specific requirements of different products within the same domain.

The prevailing paradigm for implementing an SPL is Feature-Oriented SPL (c.f. Figure 1) [5]. Features serve as focal points in communication between customers and developers regarding the product. They highlight the commonalities and differences among derived end products. During development, two categories are distinguished: Domain Engineering and Application Engineering. While the Domain Engineering focus on the development of the common platform, its features and artefacts, the application engineering focuses on the generation of a single product configuration based on a feature selection. Artefacts are the building blocks used to implement the features. The goal hereby is to reuse as much as possible.

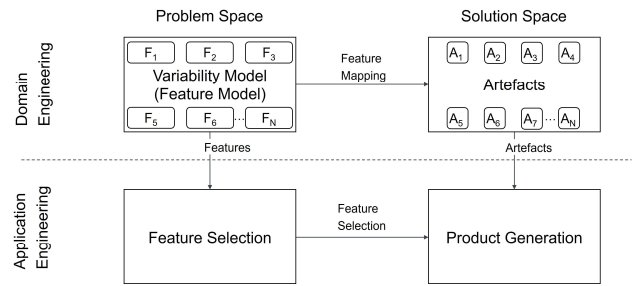

Fig. 1: Feature-Oriented Software Product Line Engineering

*2.2. Evolution of Software Product Lines*

Evolving software product lines presents a significant challenge due to their inherent complexity and long-term nature compared to traditional software systems. There are numerous possible variants and dependencies among individual software artifacts. Furthermore, programming and documentation approaches may change over the product lifecycle. Additionally, it is crucial to ensure that a software update does not render an existing (old) customer variant unusable. Developing an update is much more resource-intensive than with conventional software. Changes must be consistently implemented across all models and in the software itself to prevent malfunctions and miscommunication.

When evolving such complex systems, it inevitably leads to erosion over time [6]. This is a gradual process that is not always immediately recognized due to the complexity of the system and causes an impairment of the sustainability of the system, possibly resulting in errors or system failures. Many studies have been published regarding the erosion phenomenon of single product systems [7–10].

However, almost no research focuses on erosion in VRS and its added complexity. A practice of software product line maintenance and evolution is described by Jiang et al. [11] focusing also on mitigating design erosion. The understanding of erosion for a software product line is not further described. Michalik et al. [12] are mentioning a "systems drift" in the context of software product lines describing the deviation of the actual variant form the architectural baseline. In [13] the authors derive architecture erosion as a hindrance to effectiveness of the software product line. While these studies recognize erosion in software product lines as a relevant problem, they do not specifically define nor tackle this problem. Only two studies by Zhang et al. [14] and Cool et al. [15] are focusing directly on the aspect of erosion in VRS. [14] examine variability erosion on an industrial software product line. By deriving metrics and conducting measurements they detect potential root causes for an eroded software product line on code level. In [15] a methodology is proposed for a long-term manageable software product line architecture. The approach is based on [16] and aims at minimizing erosion of the software product line on an architectural-level.

## 3. Challenges

Facing the erosion phenomenon in variant-rich software systems unveils challenges that need to be addressed. While the concept of erosion is not new to the software engineering domain, the introduced variability in VRS adds a new layer of complexity that needs to be considered. Especially the impact of the evolution on multiple product configurations instead of only a single product needs to be investigated more closely.

Only a few studies consider erosion in relation to VRS (cf. Section 2.2). Most of them refer to erosion on the architectural level and the term "SPL erosion" is used [17, 18]. On the other hand, eroded variability elements on code level are considered or there is no further elaboration on what exactly can be understood by erosion in VRS. All in all, there is no consent nor holistic definition of erosion in VRS which hinders the discussion and research on a shared foundation.

> **C1**: Defining erosion in variant-rich software systems for a uniform understanding.

To effectively control the erosion problem in VRS, erosion must be made visible and thus measurable, which is achieved through the use of metrics. These metrics should serve the purpose of not only detecting the presence of erosion but also quantifying its extend and impact on the system. One key aspect of erosion metrics is to describe the delta between the eroded state and an "idealized", ground truth system. This allows for a clear understanding of how far the system has deviated from its intended design. Metrics should also capture the unique characteristics of VRS that are indicative of erosion. This can offer not only insights into the presence and severity of erosion within the system but might also allow an early recognition of possible erosion drivers. An effective set of metrics should enable a comprehensive evaluation of erosion within VRS and provide a holistic understanding in these complex systems. Although numerous metrics exist for quantifying erosion in software systems, almost none of them have been tailored specifically to accommodate the unique characteristics of variant-rich software [19].

> **C2**: Adapting erosion metrics for variant-rich software systems to effectively quantify and measure erosion.

Erosion in VRS can be a gradual process throughout multiple evolution steps leading to impairment of productivity and reusability of artifacts. Long-term effects (symptoms) of erosion are often only apparent in the advanced stages. Thus, an early detection of erosion, especially its root causes is a pivotal step towards tackling erosion in VRS. This can also help to prevent further erosion in future evolution steps. Although existing studies concentrate on erosion detection [20], none of them considers VRS.

> **C3**: Detecting erosion of variant-rich software systems in an early stage.

Addressing detected and ideally evaluated erosion is a critical imperative, particularly when erosion begins to impact system performance and efficiency. Therefore, repairing strategies are needed encompassing a range of actions, including refactoring, re-engineering, and error correction, all aimed at preserving the desired systems performance. However, it is important to recognize that any repair approach must consider the spectrum of system variants. It must be ensured, that these strategies do not mistakeably introduce negative effects on specific variants. Furthermore, the concept of reusability plays and important role in the context of VRS. Repair approaches should prioritize achieving high levels of artifact reusability.

> **C4**: Repairing eroded variant-rich software systems while maintaining performance and preventing negative variant impacts.

Evaluating approaches against benchmarks is key to showcase the outcome of the research. However, there is a lack in benchmarks for variant-rich software systems, especially from industry that involve a certain size and complexity. This presents a challenging task, since companies usually do not want to provide any insights into their SPL, especially not into an SPL that may be eroding in some way.

> **C5**: Benchmarking the approaches to tackle erosion in variant-rich software systems.

## 4. First Approach for Tackling Erosion in Variant-Rich Software Systems

In response to the derived challenges, we have developed an initial approach for tackling erosion in variant-rich software systems (c.f. Figure 2). Our approach centers on a common understanding of erosion in VRS. This is achieved through the conception of specific Erosion Scenarios (ES). These scenarios outline possible results of the evolution of VRS, leading to different forms of erosion. Such outcomes might pertain to erosion occurring at the architectural level or manifesting as code-level erosion. Based on the common understanding we can map expressive metrics to each ES allowing a measurement and unified description of erosion in VRS. This includes a comprehensive examination of exiting metrics and to what extent they can be used in the context of VRS.

Various detection approaches can then be applied to a VRS, analyzing the system for potential ES and to pinpoint possible erosion drivers. These approaches may leverage erosion metrics to quantitatively evaluate the extent of erosion within the system. This not only aids in early erosion detection but also provides valuable insights into the factors contributing to erosion in VRS, facilitating more targeted mitigation strategies.

Based on the detected and quantified erosion, an evaluation of possible repairing steps can be conducted. First, an evaluation is needed if any repairing is necessary based on the degree of erosion. Second, different repairing possibilities are deduced and evaluated based on their outcome. Hereby, incorporating machine learning for automated re-engineering steps leverages the effectiveness of the approach. As the majority of the data involved in the process is text-based, large language models have the potential to be a promising algorithm for assisting in the detection and repair process. In a last step the chosen repairing approach is being applied.

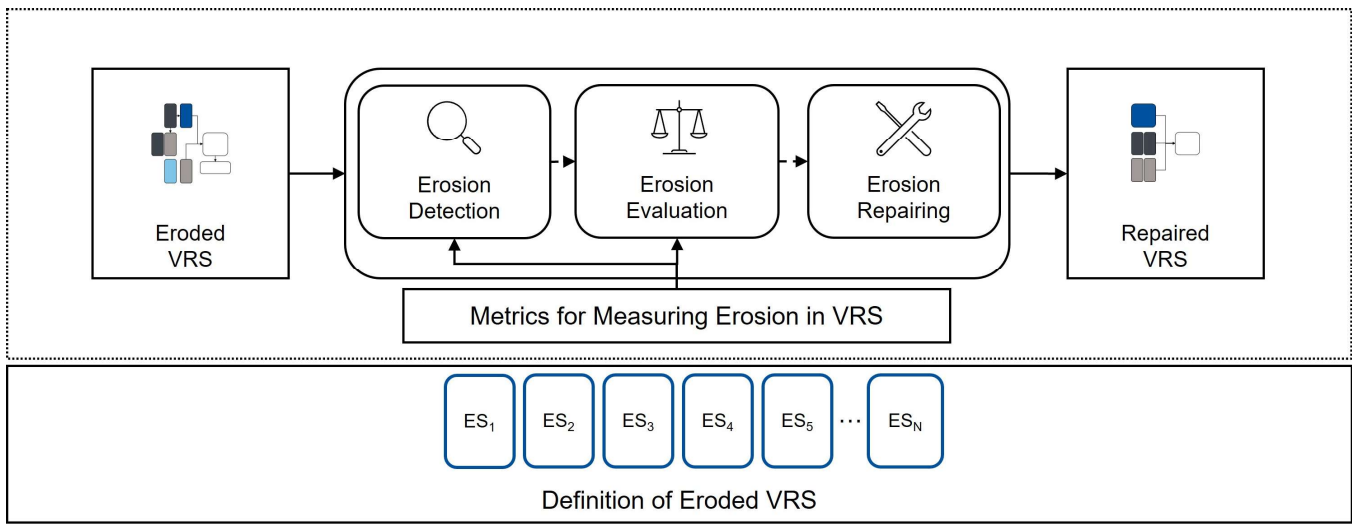

Fig. 2: First Approach for Tackling Erosion in Variant-Rich Software Systems

These steps also require a thorough evaluation of existing methods for detecting and repairing erosion, with consideration for their applicability to variant-rich software systems.

## 5. Conclusion and Future Work

In this paper we presented an overview of the state of research. Above all we derived five challenges (C1-C5) regarding tackling erosion in VRS. Especially the common understanding of erosion in VRS as well as an extensive comparison with single product systems are a crucial step towards successful research in this field. Based on the derived challenges we then presented a first high-level approach for tackling erosion in VRS. The steps involved in the approach are intended above all to consider the variability of the system.

The outcome of this paper should not only support research on tackling erosion in VRS but can also function as a basis for the development of new evolution approaches.

In future work we are focusing on the definition of the erosion scenarios as well as elaborating our first approach and refining the sub steps. Furthermore, the approach will be evaluated on a practical use case from the automotive domain.

## Acknowledgements

This publication is based on the research project SofDCar (19S21002), which is funded by the German Federal Ministry for Economic Affairs and Climate Action.